# Parametrization of 3×3 unitary matrices based on polarization algebra

**José J. Gil**

*Universidad de Zaragoza. Pedro Cerbuna 12, 50009 Zaragoza Spain*
*ppgil@unizar.es*

**Abstract**

A parametrization of 3×3 unitary matrices is presented. This mathematical approach is inspired by polarization algebra and is formulated through the identification of a set of three orthonormal three-dimensional Jones vectors representing respective pure polarization states. This approach leads to the representation of a 3×3 unitary matrix as an orthogonal similarity transformation of a particular type of unitary matrix that depends on six independent parameters, while the remaining three parameters correspond to the orthogonal matrix of the said transformation. The results obtained are applied to determine the structure of the second component of the characteristic decomposition of a 3×3 positive semidefinite Hermitian matrix.

**1 Introduction**

In many branches of Mathematics, Physics and Engineering, 3×3 unitary matrices appear as key elements for solving a great variety of problems, and therefore, appropriate parameterizations in terms of minimum sets of nine independent parameters are required for the corresponding mathematical treatment. In this way, some interesting parametrizations have been obtained [1-8]. In particular, the Cabibbo-Kobayashi-Maskawa matrix (CKM matrix) [6,7], which represents information on the strength of flavour-changing weak decays and depends on four parameters, constitutes the core of a family of parametrizations of a 3×3 unitary matrix [8]. In this paper, a new general parametrization is presented, which is inspired by polarization algebra [9] through the structure of orthonormal sets of three-dimensional Jones vectors [10]. The nine parameters of the new approach are defined by means of an orthogonal similarity transformation (thus depending on three angles) of a particular type of unitary matrix $\mathbf{V}_1$ that depends on six parameters. From the geometrical and physical point of view, such transformation can be associated with a change of the Cartesian reference frame in the three-dimensional real space, so that, unlike other known parametrizations, the one presented provides an interpretation of any given unitary matrix as a rotation transformation of the simplified unitary matrix $\mathbf{V}_1$. In particular, as we shall show later in this paper, this procedure provides a very simple physical and mathematical interpretation of sets of orthonormal vectors representing three-dimensional orthogonal states of polarization.

As an example of successful application, it is shown that the new parametrization leads to a simple analytic determination of the eigenvalues of matrices of the form $\mathrm{Re}\left[\mathbf{U}\,\mathrm{diag}(1,1,0)\,\mathbf{U}^{\dagger}\right]$ (where Re indicates "real part"; $\mathbf{U}$ is a general unitary matrix; $\mathrm{diag}(a,b,c)$ denotes a diagonal matrix with diagonal elements $(a,b,c)$, and the superscript † stands for complex conjugate). The interest in this kind of matrices comes from the fact that it has been demonstrated that they play a key role in the characterization and physical interpretation of the recently discovered nonregular states of polarization [11], with important consequences that even reach the very concept of degree of polarization of electromagnetic waves [12].

The paper is organized as follows. In Sec. 2, the three-dimensional (3D) Jones vectors and their intrinsic representation are considered as a previous step to tackle with the definition of a canonical basis (Sec. 3) which leads to a generic basis of 3D unit complex vectors through an arbitrary rotation of the Cartesian reference frame (Sec. 4), so that the said vectors can be arranged as the columns of a generic 3×3 unitary matrix, providing the desired new parametrization. The inverse problem is dealt with in Sec. 5, where the procedure for the analytic calculation of the nine parameters associated with a given unitary matrix is described in detail. As an interesting application of the approach presented, the structure of the second component of the characteristic (or trivial) decomposition [10,13] of a 3×3 positive semidefinite Hermitian matrix **R** (which determines the regularity of **R** [11]) is analyzed in Sec. 6 in the light of the new parametrization.

**2. 3D Jones vector**

Any three-component complex vector can be considered as a three-dimensional (3D) Jones vector **ε** that represents the state of polarization of fully polarized polarization state at a point in space. Such state is characterized by an ellipse (the end-point of the electric field vector of the electromagnetic wave describes the so-called polarization ellipse [9], see Fig. 1). The conventional two-dimensional (2D) model is easily reproduced by taking the $Z$ reference axis in a direction orthogonal to the plane $\Pi$ that contains the polarization ellipse, so that the third component of the 3D Jones vector vanishes.

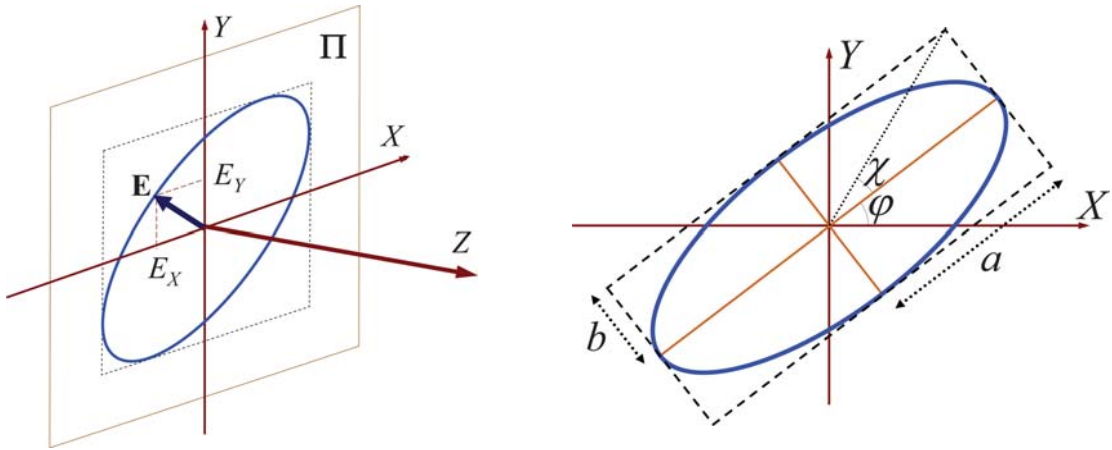

**Figure 1**. In the conventional 2D representation of polarized light the end point of the electric field vector **E** of the electromagnetic wave evolves in a fixed plane $\Pi$ so that there is no component of **E** along the $Z$ direction orthogonal to plane $\Pi$. The azimuth $\varphi$ is the angle between the direction of the major semiaxis and the $X$ axis. The absolute value of the ellipticity angle $\chi$ is given by $|\chi| = \arctan(b/a)$, $a$, $b$ being the major and minor semiaxes respectively, while the handedness is determined by the sign of $\chi$. The 2D representation does not apply to the general case in which **E** has three nonzero components for any possible Cartesian reference frame $XYZ$

Let us denote as $(\mathbf{e}_x, \mathbf{e}_y, \mathbf{e}_z)$ and $(\mathbf{e}_{x_O}, \mathbf{e}_{y_O}, \mathbf{e}_{z_O})$ the respective sets of orthonormal unit vectors characterizing an arbitrary reference frame $XYZ$ and the *intrinsic reference frame* $X_O Y_O Z_O$ of the state **ε** whose $X_O Y_O$ plane contains the polarization ellipse and whose $X_O$ axis lies in the same direction as that of the major semiaxis of the polarization ellipse of **ε** (Fig. 2). The generic expression of a given state with respect to its intrinsic reference frame $X_O Y_O Z_O$ only depends on a global phase $\gamma$, on the intensity $I$, and on the value of the ellipticity angle $\chi$ (with $-\pi/4 \leq \chi \leq \pi/4$) determining the ellipticity and the handedness of the state

$$\boldsymbol{\varepsilon}_O = \sqrt{I}\, e^{i\gamma} \begin{pmatrix} \cos\chi \\ i\sin\chi \\ 0 \end{pmatrix} \qquad (1)$$

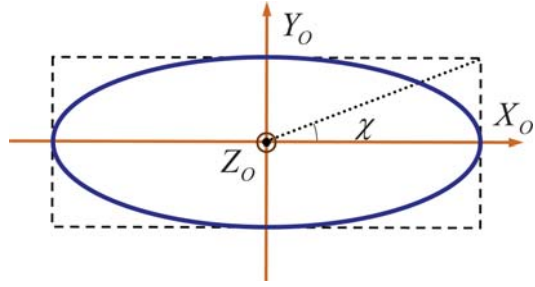

**Figure 2**. The polarization ellipse represented with respect to the intrinsic reference frame $X_O Y_O Z_O$ of a pure polarization state $\boldsymbol{\varepsilon}_O$.

Since we are interested in building a generic 3×3 unitary matrix constituted by three orthonormal vectors, hereafter we will take $I = 1$ and denote the unit Jones vector as $\hat{\boldsymbol{\varepsilon}}$

The expression for the same state $\boldsymbol{\varepsilon}_O$ when it is represented with respect to an arbitrary Cartesian reference frame *XYZ*, is obtained through the corresponding rotation from $X_O Y_O Z_O$. Such rotation is determined by 1) a rotation of angle $\varphi$ (azimuth of the polarization ellipse, $0 \le \varphi < \pi$) about the $Z_O$ axis, 2) a rotation of angle $\theta$ (elevation of the new *Z* axis, $-\pi/2 \le \theta \le \pi/2$) about the transformed *Y* axis, and 3) a rotation of angle $\phi$ (azimuth of the new *X* axis, $-\pi < \phi \le \pi$) about the transformed *Z* axis. The corresponding rotation matrices are the following

$$\mathbf{Q}_\varphi \equiv \begin{pmatrix} \cos\varphi & -\sin\varphi & 0 \\ \sin\varphi & \cos\varphi & 0 \\ 0 & 0 & 1 \end{pmatrix}, \quad \mathbf{Q}_\theta \equiv \begin{pmatrix} \cos\theta & 0 & -\sin\theta \\ 0 & 1 & 0 \\ \sin\theta & 0 & \cos\theta \end{pmatrix}, \quad \mathbf{Q}_\phi \equiv \begin{pmatrix} \cos\phi & -\sin\phi & 0 \\ \sin\phi & \cos\phi & 0 \\ 0 & 0 & 1 \end{pmatrix} \qquad (2)$$

so that the orthogonal matrix $\mathbf{Q}$ of the composed rotation is given by

$$\mathbf{Q} \equiv \mathbf{Q}_\phi \mathbf{Q}_\theta \mathbf{Q}_\varphi =$$

$$= \begin{pmatrix} \cos\phi\cos\theta\cos\varphi + \sin\phi\sin\varphi & -\cos\phi\cos\theta\sin\varphi + \sin\phi\cos\varphi & \sin\theta\cos\phi \\ -\sin\phi\cos\theta\cos\varphi + \cos\phi\sin\varphi & \sin\phi\cos\theta\sin\varphi + \cos\phi\cos\varphi & -\sin\phi\sin\theta \\ -\sin\theta\cos\varphi & \sin\theta\sin\varphi & \cos\theta \end{pmatrix}. \qquad (3)$$

Therefore, when represented with respect to an arbitrary reference frame *XYZ*, the Jones vector $\hat{\boldsymbol{\varepsilon}}$ of a given pure state adopts the form

$$\hat{\boldsymbol{\varepsilon}} = \mathbf{Q}\hat{\boldsymbol{\varepsilon}}_O = \mathbf{Q} e^{i\gamma} \begin{pmatrix} \cos\chi \\ i\sin\chi \\ 0 \end{pmatrix} =$$

$$= e^{i\gamma} \begin{pmatrix} \cos\chi(\sin\phi\sin\varphi + \cos\theta\cos\phi\cos\varphi) + i\sin\chi(\sin\phi\cos\varphi - \cos\theta\cos\phi\sin\varphi) \\ \cos\chi(\cos\phi\sin\varphi - \cos\theta\sin\phi\cos\varphi) + i\sin\chi(\cos\phi\cos\varphi + \cos\theta\sin\phi\sin\varphi) \\ -\sin\theta\cos\chi\cos\varphi + i\sin\theta\sin\chi\sin\varphi \end{pmatrix} \qquad (4)$$

The unit vector $\mathbf{e}_z(\theta,\phi)$ that gives the direction of axis *Z* with respect to the intrinsic reference frame $X_O Y_O Z_O$ is determined by the pair of angles $\theta,\phi$.

The conventions taken here about the angles allow recovering the most common expressions for the particular angular configurations and interpreting them in a natural way (note that similar but slightly different conventions are used in some related works [9,14]).

## 3. Canonical basis of 3D polarization states

Let us consider the unit Jones vector of an arbitrary pure state with intensity equal to 1, given by Eq. (4). As seen in Section 1, this pure state, when represented with respect its corresponding intrinsic reference frame $X_1Y_1Z_1$ (i.e., $\varphi = \phi = \theta = 0$) takes the form

$$\hat{\mathbf{\eta}}_1 = e^{i\gamma_1} \begin{pmatrix} \cos\chi \\ i\sin\chi \\ 0 \end{pmatrix} \tag{5}$$

From the common 2D representation of pure states, it is well known that the Jones vector $\hat{\mathbf{\eta}}_2$ whose polarization ellipse lies in the same plane $X_1Y_1$ (i.e., with $\theta_2 = \phi_2 = 0$), with azimuth $\varphi_2 = \pi/2$ and with ellipticity angle $\chi_2 = -\chi$, is orthogonal to $\hat{\mathbf{\eta}}_1$ and has the form

$$\hat{\mathbf{\eta}}_2 = e^{i\gamma_2} \begin{pmatrix} i\sin\chi \\ \cos\chi \\ 0 \end{pmatrix} \tag{6}$$

Note that in the common 2D representation of 2D states (2D states whose polarization ellipses lie in a fixed plane $X_1Y_1$) $\hat{\mathbf{\eta}}_1$ and $\hat{\mathbf{\eta}}_2$ are antipodal points on the surface of the Poincaré sphere [9].

Given the pair $\hat{\mathbf{\eta}}_1$, $\hat{\mathbf{\eta}}_2$, the set of three orthonormal Jones vector is completed with a linearly polarized pure state whose electric field vibrates along the $Z_O$ direction, whose 3D Jones vector is

$$\hat{\mathbf{\eta}}_3 = e^{i\gamma_3} \begin{pmatrix} 0 \\ 0 \\ 1 \end{pmatrix} \tag{7}$$

The states corresponding to the respective orthonormal Jones vectors are said to be orthonormal states. To avoid misunderstanding, it should be stressed that, as occurs in the common 2D representations, this concept of orthogonality of states does no imply necessarily orthogonality either between the polarization planes containing the respective polarization ellipses or between the respective propagation directions.

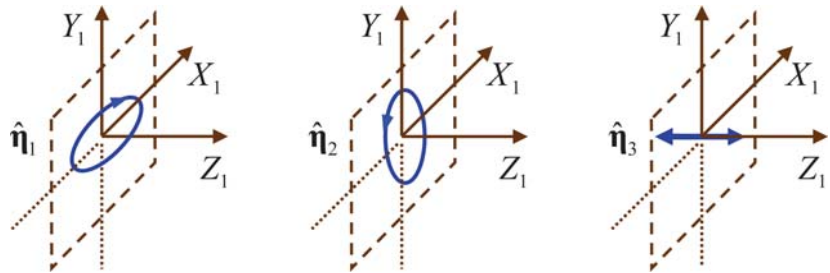

**Figure 3**. *Canonical set* of orthonormal polarization states, represented with respect to the intrinsic reference frame $X_1Y_1Z_1$ of $\hat{\mathbf{\eta}}_1$ $(\theta_1 = \Phi_1 = \varphi_1 = 0)$. State $\hat{\mathbf{\eta}}_2$ corresponds to either of the configurations $(\theta_2 = \Phi_2 = 0, \varphi_2 = \pi/2, \chi_2 = -\chi)$, $(\theta_2 = 0, \Phi_2 = \pi/2, \varphi_2 = 0, \chi_2 = -\chi)$, while $\hat{\mathbf{\eta}}_3$ corresponds to the configuration $(\theta_3 = \pm\pi/2, \varphi_3 = 0, \chi_3 = 0)$, with $\Phi_3$ arbitrary.

The *canonical* set $(\hat{\mathbf{\eta}}_1, \hat{\mathbf{\eta}}_2, \hat{\mathbf{\eta}}_3)$, constitutes an orthonormal basis for 3D Jones vectors. In general, given a polarization matrix **R**, it is always possible to determine the intrinsic reference frame $X_1Y_1Z_1$ corresponding to the first eigenvector $\hat{\mathbf{\eta}}_1$ and use the canonical basis defined with respect to it (Fig. 3).

## 4. Parametrization of a unitary matrix

Once the canonical basis has been defined, we are ready to raise the problem of identifying arbitrary sets of orthonormal polarization states.

Without loss of generality, let us take $\hat{\mathbf{\eta}}_1$ as the first vector of the basis so that it fixes the reference frame $X_1 Y_1 Z_1$ used in the equations below.

Any Jones vector $\hat{\mathbf{v}}_2$ orthonormal to $\hat{\mathbf{\eta}}_1$ can be expressed as a linear combination (with complex coefficients) of the two remaining elements of the canonical basis, that is

$$\hat{\mathbf{v}}_2 = \cos\mu \, e^{i\alpha_i} \hat{\mathbf{\eta}}_2 + \sin\mu \, e^{i\beta_i} \hat{\mathbf{\eta}}_3 = \cos\mu \, e^{i\alpha_2} \begin{pmatrix} i\sin\chi \\ \cos\chi \\ 0 \end{pmatrix} + \sin\mu \, e^{i\beta_2} \begin{pmatrix} 0 \\ 0 \\ 1 \end{pmatrix} = \begin{pmatrix} i e^{i\alpha_2} \cos\mu \sin\chi \\ e^{i\alpha_2} \cos\mu \cos\chi \\ e^{i\beta_2} \sin\mu \end{pmatrix} \quad (8)$$

with $0 \leq \mu \leq \pi/2$, $0 \leq \alpha_2 \leq \pi$, $0 \leq \beta_2 \leq \pi$.

Let $\alpha_2, \beta_2$ be an arbitrary pair of angular parameters fixing the unit vector $\hat{\mathbf{v}}_2$. Note that a generic unit vector $\hat{\mathbf{v}}_3$ orthogonal to $\hat{\mathbf{\eta}}_1$ and $\hat{\mathbf{v}}_2$ can be expressed as

$$\hat{\mathbf{v}}_3 = \begin{pmatrix} i e^{i\alpha_3} \sin\mu \sin\chi \\ e^{i\alpha_3} \sin\mu \cos\chi \\ -e^{i(\beta_2 - \alpha_2 + \alpha_3)} \cos\mu \end{pmatrix}, \quad (9)$$

with $0 \leq \mu \leq \pi/2$, $0 \leq \alpha_3 \leq \pi$, $0 \leq \beta_3 \leq \pi$.

By taking the orthonormal vectors $\hat{\mathbf{\eta}}_1, \hat{\mathbf{v}}_2, \hat{\mathbf{v}}_3$ as the columns of a unitary matrix, we find that, when represented with respect to the intrinsic reference frame of its first column vector, any unitary matrix can be written in the form

$$\mathbf{V}_1 = \begin{pmatrix} e^{i\alpha_1} \cos\chi & i e^{i\alpha_2} \cos\mu \sin\chi & i e^{i\alpha_3} \sin\mu \sin\chi \\ i e^{i\alpha_1} \sin\chi & e^{i\alpha_2} \cos\mu \cos\chi & e^{i\alpha_3} \sin\mu \cos\chi \\ 0 & i e^{i\beta_2} \sin\mu & -i e^{i(\beta_2 - \alpha_2 + \alpha_3)} \cos\mu \end{pmatrix}, \quad (10)$$

$\mathbf{V}_1$ depending on the six independent parameters $\chi, \mu, \gamma, \alpha_2, \alpha_3, \beta_2$ (two angles and four phases).

Now, by transforming the unitary matrix $\mathbf{V}_1$ in order to represent its column vectors with respect to an arbitrary reference frame *XYZ*, we get that any 3×3 unitary matrix $\mathbf{U}$ can always be expressed as follows in terms of nine parameters (five angles and four phases)

$$\mathbf{U}(\phi, \theta, \varphi, \chi, \mu, \alpha_1, \alpha_2, \alpha_3, \beta_2) =$$

$$= \mathbf{Q} \begin{pmatrix} e^{i\alpha_1} \cos\chi & i e^{i\alpha_2} \cos\mu \sin\chi & i e^{i\alpha_3} \sin\mu \sin\chi \\ i e^{i\alpha_1} \sin\chi & e^{i\alpha_2} \cos\mu \cos\chi & e^{i\alpha_3} \sin\mu \cos\chi \\ 0 & e^{i\beta_2} \sin\mu & -e^{i(\beta_2 - \alpha_2 + \alpha_3)} \cos\mu \end{pmatrix} \mathbf{Q}^T \quad (11)$$

$$\begin{bmatrix} -\pi < \phi \leq \pi, \ -\pi/2 \leq \theta \leq \pi/2, \ 0 \leq \varphi < \pi, \ -\pi/4 \leq \chi \leq \pi/4 \\ 0 \leq \mu \leq \pi/2, \ 0 \leq \alpha_1, \alpha_2, \alpha_3, \beta_2 \leq \pi \end{bmatrix}$$

where $\mathbf{Q}$ is the proper orthogonal matrix $(\mathbf{Q} = \mathbf{Q}^T, \det \mathbf{Q} = +1)$ (the superscript *T* denotes transpose)

$$\mathbf{Q} \equiv \mathbf{Q}_\phi \mathbf{Q}_\theta \mathbf{Q}_\varphi = \begin{pmatrix} \cos\phi\cos\theta\cos\varphi + \sin\phi\sin\varphi & -\cos\phi\cos\theta\sin\varphi + \sin\phi\cos\varphi & \sin\theta\cos\phi \\ -\sin\phi\cos\theta\cos\varphi + \cos\phi\sin\varphi & \sin\phi\cos\theta\sin\varphi + \cos\phi\cos\varphi & -\sin\phi\sin\theta \\ -\sin\theta\cos\varphi & \sin\theta\sin\varphi & \cos\theta \end{pmatrix} \quad (12)$$

Obviously, five additional similar parametrizations are achieved by changing the order of the columns in Eq. (11).

## 5. Obtainment of the parameters associated with a given unitary matrix

In this section, we address the problem of obtaining the angular parameters that correspond to a given unitary matrix $\mathbf{U}$. Let us denote by $\hat{\mathbf{u}}_1, \hat{\mathbf{u}}_2, \hat{\mathbf{u}}_3$ the unit orthonormal vectors that constitute the columns of $\mathbf{U}$. The procedure for the calculation of the parameters is composed of the following subsequent steps.

Let us take $\hat{\mathbf{u}}_1 \equiv (u_{11}, u_{21}, u_{31})^T$ and denote the real and imaginary parts of its three components as follows

$$a_1 \equiv \mathrm{Re}(u_{11}),\ b_1 \equiv \mathrm{Im}(u_{11}),\ a_2 \equiv \mathrm{Re}(u_{21}),\ b_2 \equiv \mathrm{Im}(u_{21}),\ a_3 \equiv \mathrm{Re}(u_{31}),\ b_3 \equiv \mathrm{Im}(u_{31}). \quad (13)$$

The generic expression for $\hat{\mathbf{u}}_1$ is

$$\hat{\mathbf{u}}_1 = e^{i\gamma} \begin{pmatrix} c_\chi (s_\phi s_\varphi + c_\theta c_\phi c_\varphi) + i s_\chi (s_\phi c_\varphi - c_\theta c_\phi s_\varphi) \\ c_\chi (c_\phi s_\varphi - c_\theta s_\phi c_\varphi) + i s_\chi (c_\phi c_\varphi + c_\theta s_\phi s_\varphi) \\ -c_\chi s_\theta c_\varphi + i s_\chi s_\theta c_\varphi \end{pmatrix}. \quad (14)$$

The unknowns are the ellipticity angle $\chi$ of the pure state $\hat{\mathbf{u}}_1$, the angular parameters $\varphi, \theta, \phi$ of the rotation matrix $\mathbf{Q}^T$ that transforms $\hat{\mathbf{u}}_1$ in $\hat{\boldsymbol{\eta}}_1$, the angle $\mu$ and the phases $\alpha_1, \alpha_2, \alpha_3, \beta_2$

$$\mathbf{Q}^T = \begin{pmatrix} c_\phi c_\theta c_\varphi + s_\phi s_\varphi & -s_\phi c_\theta c_\varphi + c_\phi s_\varphi & -s_\theta c_\varphi \\ -c_\phi c_\theta s_\varphi + s_\phi c_\varphi & s_\phi c_\theta s_\varphi + c_\phi c_\varphi & s_\theta s_\varphi \\ c_\phi s_\theta & -s_\phi s_\theta & c_\theta \end{pmatrix}, \quad \mathbf{Q}^T \hat{\mathbf{u}}_1 = \hat{\boldsymbol{\eta}}_1 = e^{i\alpha_1} \begin{pmatrix} \cos\chi \\ i\sin\chi \\ 0 \end{pmatrix}. \quad (15)$$

To calculate the set of parameters it is necessary to solve the following set of trigonometric equations, where the real and imaginary parts of the components of $\hat{\mathbf{u}}_1$ are considered as known data

$$\begin{aligned}
a_1 &\equiv \mathrm{Re}(u_{11}) = \cos\chi (\sin\phi\sin\varphi + \cos\theta\cos\phi\cos\varphi) \\
b_1 &\equiv \mathrm{Im}(u_{11}) = \sin\chi (\sin\phi\cos\varphi - \cos\theta\cos\phi\sin\varphi) \\
a_2 &\equiv \mathrm{Re}(u_{21}) = \cos\chi (\cos\phi\sin\varphi - \cos\theta\sin\phi\cos\varphi) \\
b_2 &\equiv \mathrm{Im}(u_{21}) = \sin\chi (\cos\phi\cos\varphi + \cos\theta\sin\phi\sin\varphi) \\
a_3 &\equiv \mathrm{Re}(u_{31}) = -\sin\theta\cos\chi\,\cos\varphi \\
b_3 &\equiv \mathrm{Im}(u_{31}) = \sin\theta\sin\chi\,\sin\varphi
\end{aligned} \quad (16)$$

From the above equations, the parameters are obtained unambiguously through the following steps

1) The absolute value of $\chi$ is given by either of the two following equivalent equalities

$$|\cos\chi| = \sqrt{a_1^2 + a_2^2 + a_3^2}, \quad |\sin\chi| = \sqrt{b_1^2 + b_2^2 + b_3^2}. \tag{17}$$

Since $-\pi/4 \le \chi \le \pi/4$, either of the above equations determines the absolute value of $\chi$ but not its sign. The sign of $\chi$ is determined through the procedure indicated in the Appendix.

2) $\tan\varphi = -\dfrac{b_3}{a_3}\cot\chi \Rightarrow \varphi = \arctan\left[-\dfrac{b_3}{a_3}\cot\chi\right] \quad (0 \le \varphi \le \pi)$

3) $\sin\phi = \dfrac{a_1 \sin\varphi}{\cos\chi} + \dfrac{b_1 \cos\varphi}{\sin\chi}, \quad \cos\phi = \dfrac{a_2 \sin\varphi}{\cos\chi} + \dfrac{b_2 \cos\varphi}{\sin\chi}$

4) $\sin\theta = \dfrac{b_3}{\sin\chi \sin\varphi} \quad (-\pi/2 \le \theta \le \pi/2)$

5) The orthogonal matrix $\mathbf{Q}$ is determined from the parameters $\phi, \theta, \varphi$ obtained above, so that the core matrix $\mathbf{V}_1$ (whose elements are denoted as $v_{ij}$; $i,j = 1,2,3$) is obtained through $\mathbf{V}_1 = \mathbf{Q}^T \mathbf{U} \mathbf{Q}$. Then, from $\mathbf{V}_1$, the remaining parameters are given by the following equations

$$\sin\mu = \sqrt{v_{23}^* v_{23}} = \sqrt{v_{31}^* v_{31} + v_{32}^* v_{32}}, \quad \text{or, equivalently,} \quad \cos\mu = \sqrt{v_{33}^* v_{33}} = \sqrt{v_{21}^* v_{21} + v_{22}^* v_{22}}$$

(recall that $\mu$ is defined in the range $0 \le \mu \le \pi/2$)

$$\cos\alpha_1 = \dfrac{\operatorname{Re}(v_{11})}{\cos\chi}, \quad \sin\alpha_1 = \dfrac{\operatorname{Im}(v_{11})}{\cos\chi}$$

$$\left(-\pi/4 \le \chi \le \pi/4 \Rightarrow \sqrt{1/2} \le \cos\chi \le 1\right)$$

$$\cos\alpha_2 = \dfrac{\operatorname{Re}(v_{22})}{\cos\mu \cos\chi}, \quad \sin\alpha_2 = \dfrac{\operatorname{Im}(v_{22})}{\cos\mu \cos\chi} \quad (\mu \ne \pi/2)$$

(note that when $\mu = \pi/2$, $\alpha_2$ does not appear in the parametrization)

$$\cos\alpha_3 = \dfrac{\operatorname{Re}(v_{23})}{\sin\mu \cos\chi}, \quad \sin\alpha_3 = \dfrac{\operatorname{Im}(v_{23})}{\sin\mu \cos\chi} \quad (\mu \ne 0)$$

(note that when $\mu = 0$, $\alpha_3$ does not appear in the parametrization)

$$\cos\beta_2 = \dfrac{\operatorname{Re}(v_{23})}{\sin\mu}, \quad \sin\beta_2 = \dfrac{\operatorname{Im}(v_{23})}{\sin\mu} \quad (\mu \ne 0)$$

(note that when $\mu = 0$, $\beta_2$ does not appear in the parametrization)

## 6. Structure of the second component of the characteristic decomposition of a positive semidefinite Hermitian matrix

Let us consider a 3×3 positive semidefinite Hermitian matrix **R**, which can always diagonalized as

$$\mathbf{R} = \mathbf{U}\,\mathrm{diag}(\lambda_1,\lambda_2,\lambda_3)\,\mathbf{U}^\dagger = (\mathrm{tr}\,\mathbf{R})\,\mathbf{U}\,\mathrm{diag}(\hat{\lambda}_1,\hat{\lambda}_2,\hat{\lambda}_3)\,\mathbf{U}^\dagger,$$
$$\left[\hat{\lambda}_i \equiv \lambda_i/(\mathrm{tr}\,\mathbf{R}),\quad (i=1,2,3)\right], \tag{18}$$

where **U** is the unitary matrix whose columns are the eigenvectors of **R**, and $\mathrm{diag}(\lambda_1,\lambda_2,\lambda_3)$ represents the diagonal matrix composed of the nonnegative eigenvalues, taken in nonincreasing order $(\lambda_1 \geq \lambda_2 \geq \lambda_3 \geq 0)$.

As seen in previous works [10-13], a privileged view of the structure of statistical purity of **R** is given by its *characteristic* decomposition

$$\mathbf{R} = (\mathrm{tr}\,\mathbf{R})\left[P_1 \hat{\mathbf{R}}_p + (P_2 - P_1)\hat{\mathbf{R}}_m + (1 - P_2)\hat{\mathbf{R}}_{u-3D}\right],$$
$$\hat{\mathbf{R}}_p \equiv \mathbf{U}\,\mathrm{diag}(1,0,0)\,\mathbf{U}^\dagger,\quad \hat{\mathbf{R}}_m \equiv \frac{1}{2}\mathbf{U}\,\mathrm{diag}(1,1,0)\,\mathbf{U}^\dagger,\quad \hat{\mathbf{R}}_{u-3D} \equiv \frac{1}{3}\mathbf{I}, \tag{19}$$

where **I** is the 3×3 identity matrix, the dagger indicates conjugate transpose, and $(P_1, P_2)$ are the so-called *indices of polarimetric purity* [15], defined as

$$P_1 = \hat{\lambda}_1 - \hat{\lambda}_2,\qquad P_2 = \hat{\lambda}_1 + \hat{\lambda}_2 - 2\hat{\lambda}_3. \tag{20}$$

A state **R** is said to be *regular* when $\mathbf{U}\,\mathrm{diag}(1,1,0)\,\mathbf{U}^\dagger$ is a real matrix [11] and the nonregularity of **R** is a property linked to the balance between the imaginary and real parts of the middle component $\hat{\mathbf{R}}_m$, thus stressing the interest of analyzing the structure and properties of $\hat{\mathbf{R}}_m$. Note that

$$\hat{\mathbf{R}}_m = \frac{1}{2}\left[(\mathbf{u}_1 \otimes \mathbf{u}_1^\dagger) + (\mathbf{u}_2 \otimes \mathbf{u}_2^\dagger)\right] \tag{21}$$

where $\mathbf{u}_1$ and $\mathbf{u}_2$ are the eigenvectors with larger eigenvalues and $\otimes$ stands for the Kronecker product.

From the arguments dealt with in Sec. 4, there follows that it is always possible to consider the intrinsic reference frame $X_3 Y_3 Z_3$ of the third eigenvector $\mathbf{u}_3$ as the Cartesian reference frame for the set of eigenvectors that constitute the columns of the unitary matrix **U** that diagonalizes a given **R**. Thus, it is always possible to find an orthogonal matrix **Q** such that

$$\mathbf{U} = \mathbf{Q}\,\mathbf{U}_3\,\mathbf{Q}^T$$

$$\mathbf{U}_3 \equiv (\mathbf{v}_1,\mathbf{v}_2,\mathbf{v}_3) = \begin{pmatrix} i\,e^{i\alpha_2}\cos\mu\sin\chi & i\,e^{i\alpha_3}\sin\mu\sin\chi & e^{i\alpha_1}\cos\chi \\ e^{i\alpha_2}\cos\mu\cos\chi & e^{i\alpha_3}\sin\mu\cos\chi & i\,e^{i\alpha_1}\sin\chi \\ i\,e^{i\beta_2}\sin\mu & -i\,e^{i(\beta_2-\alpha_2+\alpha_3)}\cos\mu & 0 \end{pmatrix} \tag{22}$$

where the column vectors of $\mathbf{U}_3$ have been taken with the forms $\hat{\mathbf{v}}_1 = \hat{\mathbf{v}}_2$, $\hat{\mathbf{v}}_2 = \hat{\mathbf{v}}_3$, $\hat{\mathbf{v}}_3 = \hat{\boldsymbol{\eta}}_1$, ($\hat{\boldsymbol{\eta}}_1, \hat{\mathbf{v}}_2, \hat{\mathbf{v}}_3$ being the vectors defined in Sec. 4). Therefore,

$$\hat{\mathbf{R}}_m = \mathbf{Q}\,\hat{\mathbf{R}}_{m3}\,\mathbf{Q}^T$$
$$\hat{\mathbf{R}}_{m3} \equiv \frac{1}{2}\left[(\hat{\mathbf{v}}_1 \otimes \hat{\mathbf{v}}_1^\dagger) + (\hat{\mathbf{v}}_2 \otimes \hat{\mathbf{v}}_2^\dagger)\right], \tag{23}$$

with

$$\hat{\mathbf{v}}_1 \otimes \hat{\mathbf{v}}_1^\dagger = \begin{pmatrix} c_\mu^2 s_\chi^2 & i c_\mu^2 s_\chi c_\chi & e^{-i\varphi} c_\mu s_\mu s_\chi \\ -i c_\mu^2 s_\chi c_\chi & c_\mu^2 c_\chi^2 & -i e^{-i\varphi} c_\mu s_\mu c_\chi \\ e^{i\varphi} c_\mu s_\mu s_\chi & i e^{i\varphi} c_\mu s_\mu c_\chi & s_\mu^2 \end{pmatrix}$$

$$\hat{\mathbf{v}}_2 \otimes \hat{\mathbf{v}}_2^\dagger = \begin{pmatrix} s_\mu^2 s_\chi^2 & i s_\mu^2 c_\chi s_\chi & -e^{-i\varphi} c_\mu s_\mu s_\chi \\ -i s_\mu^2 c_\chi s_\chi & s_\mu^2 c_\chi^2 & i e^{-i\varphi} c_\mu s_\mu c_\chi \\ -e^{i\varphi} s_\mu s_\mu s_\chi & -i e^{i\varphi} c_\mu s_\mu c_\chi & c_\mu^2 \end{pmatrix}$$

(24)

where the condensed notation $s_\alpha \equiv \sin\alpha$, $c_\alpha \equiv \cos\alpha$ is used, and therefore,

$$\hat{\mathbf{R}}_{m3} = \frac{1}{2}\begin{pmatrix} s_\chi^2 & 0 & 0 \\ 0 & c_\chi^2 & 0 \\ 0 & 0 & 1 \end{pmatrix} + \frac{i}{2}\begin{pmatrix} 0 & c_\chi s_\chi & 0 \\ -c_\chi s_\chi & 0 & 0 \\ 0 & 0 & 0 \end{pmatrix}$$

(25)

with $-\pi/4 \leq \chi \leq \pi/4$. Thus, while the eigenvalues of $\hat{\mathbf{R}}_m$ are $(1/2, 1/2)$, the eigenvalues of $\operatorname{Re}(\hat{\mathbf{R}}_m)$ (that determine the degree of regularity of **R** [11]), taken in nonincreasing order, are given by the diagonal elements of $\hat{\mathbf{R}}_{m3}$

$$\hat{m}_1 = \frac{1}{2}, \quad \hat{m}_2 = \frac{\cos^2\chi}{2}, \quad \hat{m}_3 = \frac{\sin^2\chi}{2},$$

(26)

so that **R** is regular (in the polarimetric sense) if and only if $\chi = 0$ (i.e. $\operatorname{Im}(\hat{\mathbf{R}}_m) = \mathbf{0}$), while maximum nonregularity is achieved when $\chi = \pm\pi/4$.

The above result constitutes an interesting application example of the parametrization presented in this work.

## Acknowledgement

The author acknowledges helpful comments and suggestions by I. San José and Tero Setälä.

## Appendix

This appendix is devoted to the procedure for the determination of the sign of parameter $\chi$ from Eqs. (16), where $a_i$ and $b_i$ are considered as data.

a) When $a_i \neq 0$ and $b_i \neq 0$ $(i=1,2,3)$, their signs imply the following restrictions on the signs of $\theta$, $\varphi$ and $\chi$.

$$a_3 < 0 \Rightarrow \begin{cases} \theta > 0 \text{ and } \varphi < \pi/2 \\ \text{or} \\ \theta < 0 \text{ and } \varphi > \pi/2 \end{cases} \qquad b_3 < 0 \Rightarrow \begin{cases} \theta < 0 \text{ and } \chi > 0 \text{ and } \varphi < \pi/2 \\ \text{or} \\ \theta > 0 \text{ and } \chi < 0 \text{ and } \varphi < \pi/2 \\ \text{or} \\ \theta > 0 \text{ and } \chi > 0 \text{ and } \varphi > \pi/2 \\ \text{or} \\ \theta < 0 \text{ and } \chi < 0 \text{ and } \varphi > \pi/2 \end{cases}$$

$$a_3 > 0 \Rightarrow \begin{cases} \theta < 0 \text{ and } \varphi < \pi/2 \\ \text{or} \\ \theta > 0 \text{ and } \varphi > \pi/2 \end{cases} \qquad b_3 > 0 \Rightarrow \begin{cases} \theta > 0 \text{ and } \chi > 0 \text{ and } \varphi < \pi/2 \\ \text{or} \\ \theta > 0 \text{ and } \chi < 0 \text{ and } \varphi > \pi/2 \\ \text{or} \\ \theta < 0 \text{ and } \chi < 0 \text{ and } \varphi < \pi/2 \\ \text{or} \\ \theta < 0 \text{ and } \chi > 0 \text{ and } \varphi > \pi/2 \end{cases}$$

Therefore, in the above cases the sign of $\chi$ is determined by means of the following table:

$$\begin{aligned} a_3 < 0, b_3 < 0 &\Rightarrow \chi < 0 \\ a_3 < 0, b_3 > 0 &\Rightarrow \chi > 0 \\ a_3 > 0, b_3 < 0 &\Rightarrow \chi > 0 \\ a_3 > 0, b_3 > 0 &\Rightarrow \chi < 0 \end{aligned} \tag{27}$$

The remaining particular cases to determine the sign of $\chi$ are the following

b) When $a_3 = b_3 = 0$, we have the following possibilities

    b1) $\sin\chi = 0 \; (\chi = 0)$ and $\cos\varphi = 0 \; (\varphi = \pi/2)$, with $\theta \neq 0$

    b2) $\sin\theta = 0 \Rightarrow \theta = 0 \Rightarrow \cos\theta = +1$ (recall that $\theta$ is defined in the range $-\pi/2 \leq \theta \leq \pi/2$, which implies $\cos\theta > 0$). In this case, Eqs. (16) become the following

$$\begin{aligned} a_1 &= \cos\chi (\sin\phi \sin\varphi + \cos\phi \cos\varphi) \\ b_1 &= \sin\chi (\sin\phi \cos\varphi - \cos\phi \sin\varphi) \\ a_2 &= \cos\chi (\cos\phi \sin\varphi - \sin\phi \cos\varphi) \\ b_2 &= \sin\chi (\cos\phi \cos\varphi + \sin\phi \sin\varphi) \end{aligned} \tag{28}$$

and consequently,

$$\tan\chi = \frac{b_2}{a_1} \Rightarrow \begin{matrix} a_1 < 0, b_2 < 0 \Rightarrow \chi > 0 \\ a_1 < 0, b_2 > 0 \Rightarrow \chi < 0 \\ a_1 > 0, b_2 < 0 \Rightarrow \chi < 0 \\ a_1 > 0, b_2 > 0 \Rightarrow \chi > 0 \end{matrix} \quad \text{or, equivalently,} \quad \tan\chi = -\frac{b_1}{a_2} \Rightarrow \begin{matrix} a_2 < 0, b_1 < 0 \Rightarrow \chi < 0 \\ a_2 < 0, b_1 > 0 \Rightarrow \chi > 0 \\ a_2 > 0, b_1 < 0 \Rightarrow \chi > 0 \\ a_2 > 0, b_1 > 0 \Rightarrow \chi < 0 \end{matrix}$$

c) When $a_3 = 0$ and $b_3 \neq 0$, then $\cos\varphi = 0 \; (\varphi = \pi/2)$, with $\theta \neq 0$. In this case, Eqs. (16) become:

$$\begin{aligned} a_1 &= \cos\chi \sin\phi, & a_2 &= \cos\chi \cos\phi, & a_3 &= 0, \\ b_1 &= -\sin\chi \cos\theta \cos\phi, & b_2 &= \sin\chi \cos\theta \sin\phi, & b_3 &= \sin\theta \sin\chi, \end{aligned} \tag{29}$$

and consequently,

$$\cot\chi = \frac{a_1}{b_2}\cos\theta \Rightarrow \begin{matrix} a_1 < 0, b_2 < 0 \Rightarrow \chi > 0 \\ a_1 < 0, b_2 > 0 \Rightarrow \chi < 0 \\ a_1 > 0, b_2 < 0 \Rightarrow \chi < 0 \\ a_1 > 0, b_2 > 0 \Rightarrow \chi > 0 \end{matrix} \quad \text{or, equivalently,} \quad \cot\chi = -\frac{a_2}{b_1}\cos\theta \Rightarrow \begin{matrix} a_2 < 0, b_1 < 0 \Rightarrow \chi < 0 \\ a_2 < 0, b_1 > 0 \Rightarrow \chi > 0 \\ a_2 > 0, b_1 < 0 \Rightarrow \chi > 0 \\ a_2 > 0, b_1 > 0 \Rightarrow \chi < 0 \end{matrix}$$

d) When $a_3 \neq 0$ and $b_3 = 0$, we have the following alternative possibilities

d1) $\sin\chi = 0 \Rightarrow \chi = 0 \;(\Rightarrow b_1 = b_2 = 0)$

d2) $\sin\varphi = 0 \Rightarrow \varphi = 0, \pi$. Then, Eqs. (16) become the following

$$\begin{aligned} a_1 &= \pm\cos\chi\cos\theta\cos\phi, & a_2 &= \mp\cos\chi\cos\theta\sin\phi, & a_3 &= \mp\sin\theta\cos\chi, \\ b_1 &= \pm\sin\chi\sin\phi, & b_2 &= \pm\sin\chi\cos\phi, & b_3 &= 0, \end{aligned} \qquad (30)$$

and the equations become:

$$\tan\chi = \frac{b_2}{a_1}\cos\theta \Rightarrow \begin{array}{l} a_1<0, b_2<0 \Rightarrow \chi>0 \\ a_1<0, b_2>0 \Rightarrow \chi<0 \\ a_1>0, b_2<0 \Rightarrow \chi<0 \\ a_1>0, b_2>0 \Rightarrow \chi>0 \end{array} , \quad \text{or} \quad \tan\chi = -\frac{b_1}{a_2}\cos\theta \Rightarrow \begin{array}{l} a_2<0, b_1<0 \Rightarrow \chi<0 \\ a_2<0, b_1>0 \Rightarrow \chi>0 \\ a_2>0, b_1<0 \Rightarrow \chi>0 \\ a_2>0, b_1>0 \Rightarrow \chi<0 \end{array}$$